\documentclass[fleqn,usenatbib]{mnras}
\usepackage{newtxtext,newtxmath}

\usepackage[T1]{fontenc}

\DeclareRobustCommand{\VAN}[3]{#2}
\let\VANthebibliography\thebibliography
\def\thebibliography{\DeclareRobustCommand{\VAN}[3]{##3}\VANthebibliography}

\usepackage{graphicx}	
\usepackage{aligned-overset}
\usepackage{siunitx} 
\usepackage{hyperref} 
\usepackage[capitalise]{cleveref} 
\makeatletter
\usepackage{etoolbox}
\patchcmd\H@refstepcounter{\protected@edef}{\protected@xdef}{}{}
\makeatother

\newcommand{\hairsp}{\hspace{1pt}} 
\newcommand{\ie}{\hbox{i.\hairsp{}e.\hairsp{} }} 
\newcommand{\eg}{\hbox{e.\hairsp{}g.\hairsp{} }} 

\newcommand{\MUSIC}{\textsc{MUSIC}}

\newcommand{\rmin}{r_{\min}}
\newcommand{\rmax}{r_{\max}}
\newcommand{\rschw}{r_{\textrm{Schw}}}
\newcommand{\tconv}{\tau_\textrm{conv}}

\newcommand{\ekin}{\textrm{E}_\mathrm{k}}
\newcommand{\etot}{\textrm{E}_\mathrm{t}}
\newcommand{\krad}{k_\mathrm{rad}}

\newcommand{\lout}{L_\mathrm{out}}
\newcommand{\lin}{L_\mathrm{in}}

\newcommand{\mstar}{M_{\ast}}
\newcommand{\rstar}{R_{\ast}}
\newcommand{\lstar}{L_{\ast}}
\newcommand{\msun}{M_{\sun}}
\newcommand{\rsun}{R_{\sun}}
\newcommand{\lsun}{L_{\sun}}
\newcommand{\nablaad}{\nabla_{\mathrm{ad}}}
\newcommand{\Ny}{\mathrm{Ny}}

\newcommand{\bulk}{\mathrm{bulk}}
\newcommand{\bkgr}{\textrm{bkgr}}
\newcommand{\ke}{\mathrm{ke}}
\newcommand{\conv}{\mathrm{conv}}

\newcommand{\half}{\frac{1}{2}}

\renewcommand{\v}[1]{\boldsymbol{ #1 }}  
\newcommand{\vrms}{v_\textrm{rms}}
\newcommand{\vu}{\v{v}}
\newcommand{\vur}{v_r}

\newcommand{\bra}[1]{\left( #1 \right)}
\newcommand{\avg}[1]{\left< #1 \right>}
\newcommand{\abs}[1]{\left| #1 \right|} 

\DeclareDocumentCommand{\pdt}{g}{
	\IfNoValueTF{#1}
	{\partial_t }
	{\partial_t \bra{#1}}
}

\newcommand{\di}{\mathrm{d}}

\DeclareDocumentCommand{\div}{g}{
  \IfNoValueTF{#1}
    {\nabla \cdot}
    {\nabla \cdot \bra{#1}}
}

\DeclareDocumentCommand{\grad}{g}{
  \IfNoValueTF{#1}
    {\nabla }
    {\nabla  \bra{#1}}
}

\DeclareDocumentCommand{\pdr}{g}{
	\IfNoValueTF{#1}
	{\partial_r}
	{\partial_r \bra{#1}}
}
\DeclareDocumentCommand{\Tr}{g}{
  \IfNoValueTF{#1}
    {\mathrm{Tr}}
    {\mathrm{Tr}\bra{#1}}
}

\title[Radial truncation in 2D Sun-like model]{
Impact of radial truncation on global 2D hydrodynamic simulations for a Sun-like model
}

\author[D. G. Vlaykov et al.]{
D. G. Vlaykov,$^{1}$\thanks{d.vlaykov@exeter.ac.uk}
I. Baraffe,$^{1,2}$
T. Constantino,$^{1}$
T. Goffrey,$^{3}$
T. Guillet,$^{1}$
A. Le Saux,$^{1,2}$
A. Morison,$^{1}$
\newauthor
and J. Pratt$^{4}$
\\
$^{1}$ Physics and Astronomy, University of Exeter, Stocker Road, Exeter, EX4 4QL,UK\\
$^{2}$ \'Ecole Normale Sup\'erieure, Lyon, CRAL (UMR CNRS 5574), Universit\'e de Lyon, France\\
$^{3}$ Centre for Fusion, Space and Astrophysics,  Department of Physics, University of Warwick, Coventry, CV4 7AL, UK\\
$^{4}$ Department of Physics and Astronomy, Georgia State University, Atlanta, GA 30303, USA
}
\date{Accepted 2022 April 29. Received 2022 April 29; in original form 2022 February 22}

\pubyear{2022}

\begin{document}
\label{firstpage}
\pagerange{\pageref{firstpage}--\pageref{lastpage}}
\maketitle

\begin{abstract}
Stellar convection is a non-local process responsible for the transport of heat and chemical species.
It can lead to enhanced mixing through convective overshooting and excitation of internal gravity waves (IGWs) at convective boundaries.
The relationship between these processes is still not well understood and requires global hydrodynamic simulations to capture the important large-scale dynamics.
The steep stratification in stellar interiors suggests that the radial extent of such simulations can affect the convection dynamics, the IGWs in the stably stratified radiative zone, and the depth of the overshooting layer.
We investigate these effects using two-dimensional global simulations performed with the fully compressible stellar hydrodynamics code \MUSIC.
We compare eight different radial truncations of the same solar-like stellar model evolved over approximately $400$ convective turnover times.
We find that the location of the inner boundary has an insignificant effect on the convection dynamics, the convective overshooting and the travelling IGWs.
We relate this to the background conditions at the lower convective boundary which are unaffected by the truncation, as long as a significantly deep radiative layer is included in the simulation domain.
However, we find that extending the outer boundary by only a few percent of the stellar radius significantly increases the velocity and temperature perturbations in the convection zone, the overshooting depth, the power and the spectral slope of the IGWs.
The effect is related to the background conditions at the outer boundary, which are determined in essence by the hydrostatic stratification and the given luminosity.
\end{abstract}




\section {Introduction}
Stellar convection is a non-local process \citep{rieutord1995,spruit1997,brandenburg2016}, which can affect stellar structure and evolution \citep{spruit1997,gough1977,canuto1997}.
Geometrical parameters, such as the aspect ratio and vertical extent can play an important dynamical role in simulations of boundary-driven convection \citep{hurlburt1984,cossette2016,pratt2016,wagner2013}.
This work aims to further qualify this role, by studying the effects of varying the radial extent in global two-dimensional hydrodynamic  simulations of a Sun-like stellar model on the dynamics of the convective and radiative zones and the boundary between them.

Global stellar hydrodynamic simulations are computationally challenging because of the extreme dynamical ranges of length, time, and amplitude scales found in stellar interiors.
Typically, the geometrical domain is restricted radially at some point below the stellar photosphere and/or above the stellar core.
For instance in global solar simulations, \citet{rogers2006} set the radial range at $\SI{0.001}{\rsun} - \SI{0.93}{\rsun}$ and \citet{brun2011} use the radial range $\SI{0.07}{\rsun} - \SI{0.97}{\rsun}$.
When focusing on the solar convection zone typical radial ranges are $\SI{0.72}{\rsun} - \SI{0.97}{\rsun}$ \citep{augustson2015}, or when including convective boundary effects, $\SI{0.71}{\rsun} - \SI{0.96}{\rsun}$ \citep{hotta2021} or $\SI{0.61}{\rsun} - \SI{0.96}{\rsun}$ \citep{guerrero2013}.
While some authors consider the effect of boundary conditions (\eg  by  including an outer cooling layer \citealt{guerrero2013,hotta2019}),
the isolated effect of the radial truncation itself has rarely been investigated to our knowledge (see below).
However, both the inner and outer boundary layers can house important processes with both global and long-term effects.

The temperature and density stratification in the outer convective layers can be very steep, dropping by an order of magnitude over a percent of a stellar radius.
Thus a small change in radius leads to a large change in the background thermodynamic conditions.
Due to the inherent non-locality of boundary-driven convection \citep{gough1977,spruit1997,canuto1997}, this is expected to impact the global dynamics of the simulation domain.
In fact, it has been suggested that the outer solar layers are the dominant driver of the convective zone \citep{spruit1997,brandenburg2016}
through an `entropy rain' of radiatively cooled dense perturbations\footnotemark.
\footnotetext{
Note that this study does not attempt to address the viability of the entropy rain framework, because the considered simulation domain does not extend to regions where radiative cooling becomes significant.
We focus instead on isolating the effect of radial extension of the region with efficient convection.}

The lower convective boundary is also of particular interest.
Due to inertia, flows from the convective zones can penetrate some distance across the boundary into the stably-stratified radiative zone.
This process has been referred to as convective penetration, overshooting or convective boundary mixing \citep{zahn1991,hurlburt1994,brummell2002} in different contexts.
Here we follow the terminology in \citet{brummell2002} and use the term \emph{overshooting} to indicate convective flows traveling through sub-adiabatic layers.
Despite occurring in a small stellar region (by volume and mass), convective overshooting can have a quantitative impact on the stellar evolution.
In general, the dynamics at the lower convective boundary are central to a number of open questions of stellar physics relating to mixing and transport of material, angular momentum, and magnetic fields.
For instance, the enhanced mixing of material from the well-mixed cooler convective zone across the lower convective boundary can help deplete the surface lithium abundance  in pre-main-sequence and main-sequence stars \citep{baraffe2017,constantino2021}.
Convective motions along and across the boundary excite internal gravity waves (IGWs) in the stably-stratified radiative zone (RZ) \citep{press1981,goldreich1990,dintrans2005,pincon2016}.
IGWs in turn are a sensitive probe of stellar interior structure, observable in massive stars \citep{bedding2010,rogers2013,bowman2019}, and are involved in the redistribution of angular momentum \citep{rogers2013} and chemical mixing \citep{rogers2017}.
Finally, there is an active debate regarding the amplitude of the large-scale convective velocities in solar interior, the so-called \emph{solar convective conundrum}.
Different helioseismological techniques point to a bigger \citep{hanasoge2012} or smaller \citep{birch2018} discrepancy compared to the typical values obtained in numerical simulations.
Sometimes considered as part of the convective conundrum is also the problem of reproducing the solar-type of differential rotation in the convection zone, \ie fast equator and slow poles\footnotemark.
\footnotetext{Recent advances suggest that magnetohydrodynamic simulations with sufficiently high resolution can reproduce the solar type of differential rotation \citep{hotta2021}.}

The solutions to such questions require stellar hydrodynamic simulations, which can capture non-local processes in both space and scale.
While three-dimensional (3D)  simulations can be more realistic, their large computational cost makes extensive parameter studies challenging, especially over long time intervals.
The relaxation time scales may be shortened by artificially boosting the stellar luminosity and thermal diffusivity.
However, running such simulations self-consistently requires that the convection zone be adiabatic, to avoid changes in the temperature stratification due to the boosting \citep{baraffe2021}. It has also been shown that boosting can have a significant impact on the dynamics throughout the stellar interior \citep{baraffe2021,lesaux2021}.
Therefore, in this study, we do not resort to boosted simulations and consider instead two-dimensional (2D) models with realistic stellar luminosity covering a wide selection of radial truncations.

Several studies have examined the influence of the radial extent using different sets of approximations.
\Citet{hurlburt1984} performed  2D fully compressible simulations on a Cartesian grid with polytropic stratification and ideal gas law equation of state (EoS) and found that mean kinetic energy grows with the density contrast in the domain (which is analogous to the depth of the convection zone).
\Citet{cossette2016} performed 3D anelastic simulations on a Cartesian grid with polytropic stratification.
They reported that the super-adiabaticity in the boundary layer affects both the amplitude of the temperature fluctuations in the convective bulk and the horizontal length-scale of the velocity field.
Such a change in super-adiabaticity at the outer boundary occurs naturally when varying the radial extent of the simulation domain using realistic stellar stratification.
The reason is that in this case hydrostatic equilibrium dictates a sharp drop in density close to the stellar surface. This results in a decrease of the total heat capacity of the outer layers and thus the amount of luminosity that the convective heat flux can transport\footnotemark.
\footnotetext{With realistic stellar EoS the specific heat at constant pressure is not sensitive enough to the temperature to counterbalance the effect of the decreasing density.}
As a result the temperature stratification must become steeper than adiabatic, so that the radiative flux can transport the remaining luminosity.
\Citet{pratt2016} performed 2D fully compressible global simulations with a realistic EoS for a low mass star, opacity and stratification.
They studied the effects on convection and overshooting due to grid resolution and the radial truncation of the simulation domain in a low-mass pre-main sequence star, a `young sun'.
They found that the radial extent of the simulation domain and the coupling at the upper and lower convective boundaries has a pronounced effect on the dynamics of the convective zone and the depth of the overshooting layer.
Finally, \citet{hotta2019} considered both local Cartesian and global spherical simulations in 3D using a reduced speed of sound technique and a combination of linear and tabulated EoS.
They found little difference in the convective bulk when comparing the effects on the convection zone from (i) including a photospheric outer layer with resolved radiative transport and (ii) adding artificial cooling with a Gaussian profile around the top of the convection zone.
They also found that local Cartesian and global spherical simulation produce similar results, and indicate that including a photospheric layer is unlikely to help solve the solar convective conundrum.

Building on these results, we perform global 2D fully-compressible simulations of a Sun-like star (with solar mass, luminosity and metallicity) on the main sequence (at $\sim \SI{4.6}{\giga yr}$) using a realistic EoS, opacity and initial stratification.
In order to isolate the impact of the radial extent of the outer convective and inner radiative zones we do not consider the effects of rotation, magnetic fields, and cooling from atmospheric layers.
Similarly to \citet{pratt2016}, we study the impact on the convective velocity and the depth of the overshooting layer in 2D simulations.
In addition, we measure the temperature fluctuations as well as the impact on the generated IGWs and their potential feedback on the overshooting.
We highlight the role of the outer boundary layers and how they can qualitatively affect the entire convection zone and the overshooting region.
We use a main-sequence Sun-like model, which has a lower luminosity, a significantly shallower convection zone and much steeper density stratification in the radiative layers than the \citet{pratt2016} model.
To our knowledge this is the first study of its type applied to a Sun-like stellar model on the main sequence.

The paper is organised as follows.
In \cref{sec:methods} we describe the numerical methods and the simulations.
We examine and discuss the impact of the radial truncation on the convection zone dynamics,
(\cref{sec:results cz}),
the overshooting (\cref{sec:results os}) and the IGWs in the radiative zone (\cref{sec:results rad}).
We conclude with a summary of the results and a discussion of their implications.

\section{Methods} \label{sec:methods}
\subsection {Numerics: MUSIC}

We use the fully compressible time-implicit code \MUSIC.
The code is described in detail in \cite{Viallet2011,Viallet2016,Goffrey2017}.
Here, we provide a brief description of the main features used in this study.
\MUSIC ~
solves the inviscid Euler equations in the presence of thermal diffusion and external gravity,

\begin{align}
	\pdt{\rho} &= - \div{\rho \vu},  \label{eq:cont} \\
	\pdt{\rho \vu} &= - \div {\rho \vu \vu} - \grad p + \rho \v{g},  \label{eq:ns} \\
	\pdt{\rho e} &= -\div{\rho e \vu} - p \div \vu + \div{\chi \grad T}, \label{eq:eint}
\end{align}
with density $\rho$, velocity $\vu$, specific internal energy density $e$, gas pressure $p$, gravitational acceleration $\v{g}$,
and thermal conductivity $\chi$.
In the presented simulations, the thermal conductivity is dominated by radiative transfer:
\begin{align}
	\chi = \frac{16 \sigma T^3}{3 \kappa \rho},
\end{align}
with $\kappa$ the Rosseland mean opacity, and $\sigma$ the Stefan-Boltzmann constant.
We use realistic opacities and equation of state for stellar interiors.
The opacities are interpolated from the OPAL tables \citep{iglesias1996}
for solar metallicity and the equation of state is based on the OPAL EoS tables of \citet{Rogers2002},
which are appropriate for the description of solar-like interior structures.

We do not include explicit viscosity in the equations, because its value in stellar interiors is too low to dominate over numerical viscosity for  computationally tractable grid resolutions.
 Instead we rely on the numerical dissipation of the model and interpret the results in the context of the implicit large-eddy simulation (ILES) paradigm.

\subsection{Description of the simulations.} \label{sec:sims}

\begin{figure}
	\includegraphics[width=\columnwidth]{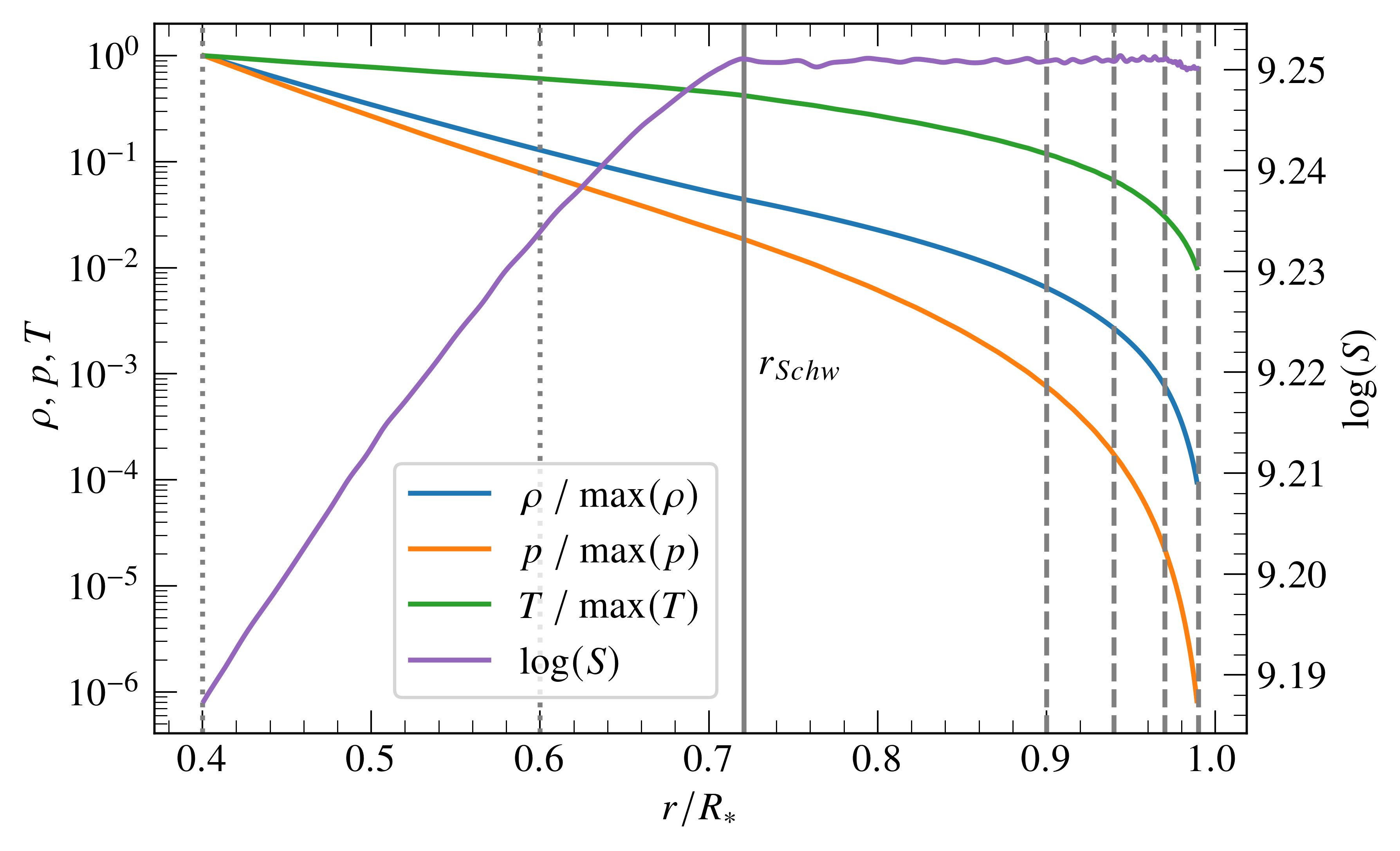} \\
	\caption{
	Initial background radial profile in density, temperature, pressure and entropy.
	The first three are normalised to their values at $0.4\rstar$ to better compare the stratification in the upper layers.
	The solid vertical line shows the boundary between the convective and radiative zone.
	The dotted vertical lines on the left show the truncation radii for $\rmin$,
	the dashed vertical lines on the right show the truncation radii for $\rmax$.
	\label{fig:IC}
	 }
\end{figure}

The initial background state is defined by the density and temperature stratifications shown in \cref{fig:IC} with $\rstar=\SI{7.0775 e10}{cm} = \SI{1.0169}{\rsun}$, $\lstar = \SI{3.792 e33}{erg. s^{-1}} = \SI{0.98772} {\lsun}$ and $\mstar = \SI{1.9891 e33}{g} = \SI{1}{\msun}$, where $\odot$ represents the solar reference values.
The initial conditions are generated with the one-dimensional Lyon stellar evolution code \cite{Baraffe1991,Baraffe1998},
using the same opacities and equation of state as implemented in \MUSIC.
Note that the lower boundary of the  convection zone in the model is at $\rschw \sim \num{0.72}  \rstar$ as defined by the Schwarzschild criterion, which gives the thermal stratification required for a convective instability. Explicitly, the
Schwarzschild criterion defines the convective boundary via the super-adiabaticity being non-negative, \ie $\nabla - \nablaad \geq 0$.
Here, $\nabla =\di \log T /  \di \log p$ is the temperature gradient and $\nablaad= \di \log T /  \di \log p |_S$ is the adiabatic temperature gradient, \ie at constant entropy $S$.
In practice in the convection zone the super-adiabaticity is very close to zero, as indicated by the quasi-isentropic profile shown in \cref{fig:IC}.
Note however that the convection zone is not precisely adiabatically stratified with super-adiabaticity reaching $O(10^{-2})$ close to the surface.
A close analogue of this initial model has been used in \cite{baraffe2021,lesaux2021} to study the effects of luminosity boosting.
This requires  minimising the super-adiabaticity in the convection zone (so that stratification is not affected by the luminosity boosting).
A comparison between the two models shows that  this leads to a significant change in the total stellar radius and the lower convective boundary and a slight increase in luminosity.

\Cref{tab:params} summarises the key parameters of the eight performed simulations.
We consider two lower boundaries ($\rmin /\rstar \in [0.4, 0.6]$) and four upper boundaries ($\rmax /\rstar \in [0.9, 0.94, 0.97, 0.99]$).
The one-dimensional profile is extruded symmetrically in the angular direction.
The simulation domain covers the full angular range of $\theta \in [\ang{0}, \ang{180}]$ with angular resolution of $\Delta \theta = \ang{0.357 }$.
We use a fixed radial resolution for all simulations of $\Delta r = \num{6 e-4} \rstar \approx \SI{420}{km}$.
The radial cell size is chosen to ensure good representation of the pressure scale height  $H_p(r)=-\partial r/\partial \ln(p(r))$ even at the outermost simulated layers.
Thus, at the outer boundaries the radial cell size corresponds to  $\Delta r = 0.225 H_p(0.99 \rstar)$ and $ \Delta r =0.017 H_p(0.9 \rstar)$.
The large range of $\Delta r /H_p(\rmax)$ is due to the strong stratification which leads the pressure scale height to decrease sharply in the top $\sim 10\%$ of the stellar radius, as can be inferred from \cref{fig:IC}.
At the lower convective boundary, the radial cell size corresponds to  $\Delta r = 0.0075H_p(\rschw)$.

At the polar boundaries ($\theta = \ang{0}$ and $\theta = \ang{180}$), the normal derivative is set to zero for all fields.
At the radial boundaries, the boundary conditions are reflective for the velocity.
The density and internal energy are linearly extrapolated in the ghost cells used to calculate the spatial gradients and fluxes.
This condition is well-suited to the stratified background.
Due to the use of staggered grids, the radiative flux at the radial boundary must also be given.
Here, it is set to ensure a constant uniform energy flux corresponding to the respective luminosity of the 1D model at that location.
In practice, this makes for a slightly different luminosity at the inner boundary between the $\rmin = 0.4 \rstar$ and  $\rmin = 0.6 \rstar$ simulations (see \cref{tab:params}) but has no measurable difference at the outer boundaries where $\lout= \SI{0.98772}{ \lsun}$ for all simulations.
The slight difference between the inner and outer boundary drives the background state to evolve, however this happens on the thermal time-scale which is much longer than the duration of the simulations, as discussed below.

\begin{table*}

\caption{
	Simulation parameters: inner  and outer simulation boundary,
	number of cells in the radial direction $N_r$,
	radial cell width in units of the pressure scale height at the outer boundary,
	total simulation duration $\tau_\mathrm{tot}$ in units of the convective turnover time,
	the steady-state convective turnover time $\tconv$,
	the inner luminosity in units of the solar luminosity,
	the average temperature at the outer boundary $T(\rmax)$,
	the average density at the outer boundary $\rho(\rmax)$,
	hydrodynamic boundary layer thickness $\delta_{\vur}$ in units of the stellar radius.
	\label{tab:params}
	}
\begin{tabular}{c l  c c c   c c c c c c}
\hline
$\rmin/ \rstar$ &
 $\rmax/\rstar$ &
$N_r$ &
$\Delta_r/H_p(\rmax)$ &
 $\tau_\mathrm{tot}/ \tconv$  &
 $\tconv/\si{10^5 s}$ &
$\lin/\lsun$ &

$T(\rmax)/\si{10^5 K}$ &
$\rho(\rmax)/\si{10^{-3} g.cm^{-3}}$ &
$\delta_{\vur}/ \rstar$ &
\\ \hline \hline
	 0.4           & 0.9            &840  & 0.017  & $346$         &   5.8   &   0.98778   & 6.10  & 23.7   & 0.010 \\
	 0.4           & 0.94          &912  & 0.030  & $374$         &   5.4   &  0.98778    & 3.44  & 9.9	   & 0.009 \\
         0.4           & 0.97          &960  & 0.060  & $547$         &   4.7   &  0.98778    & 1.60  & 2.9      & 0.007 \\
	 0.4           & 0.99          &996  & 0.220  & $615$         &   4.3   &  0.98778    & 0.51  & 0.3     & 0.002 \\ \hline
	 0.6           & 0.9            &504  & 0.017  & $366$         &   5.9   &  0.98775    & 6.10  & 23.7   & 0.010 \\
	 0.6           & 0.94          &576  & 0.030  & $730$         &   5.4   &  0.98775    & 3.44  & 9.9     & 0.009 \\
	 0.6           & 0.97          &624  & 0.061  & $551$         &   4.6   &  0.98775    & 1.60  & 2.9     & 0.007 \\
	 0.6           & 0.99          &660  & 0.220  & $585$         &   4.2   &  0.98775    & 0.51  & 0.3     & 0.002 \\ \hline
\end{tabular}
\end{table*}

\begin{figure}
	\includegraphics[width=\columnwidth]{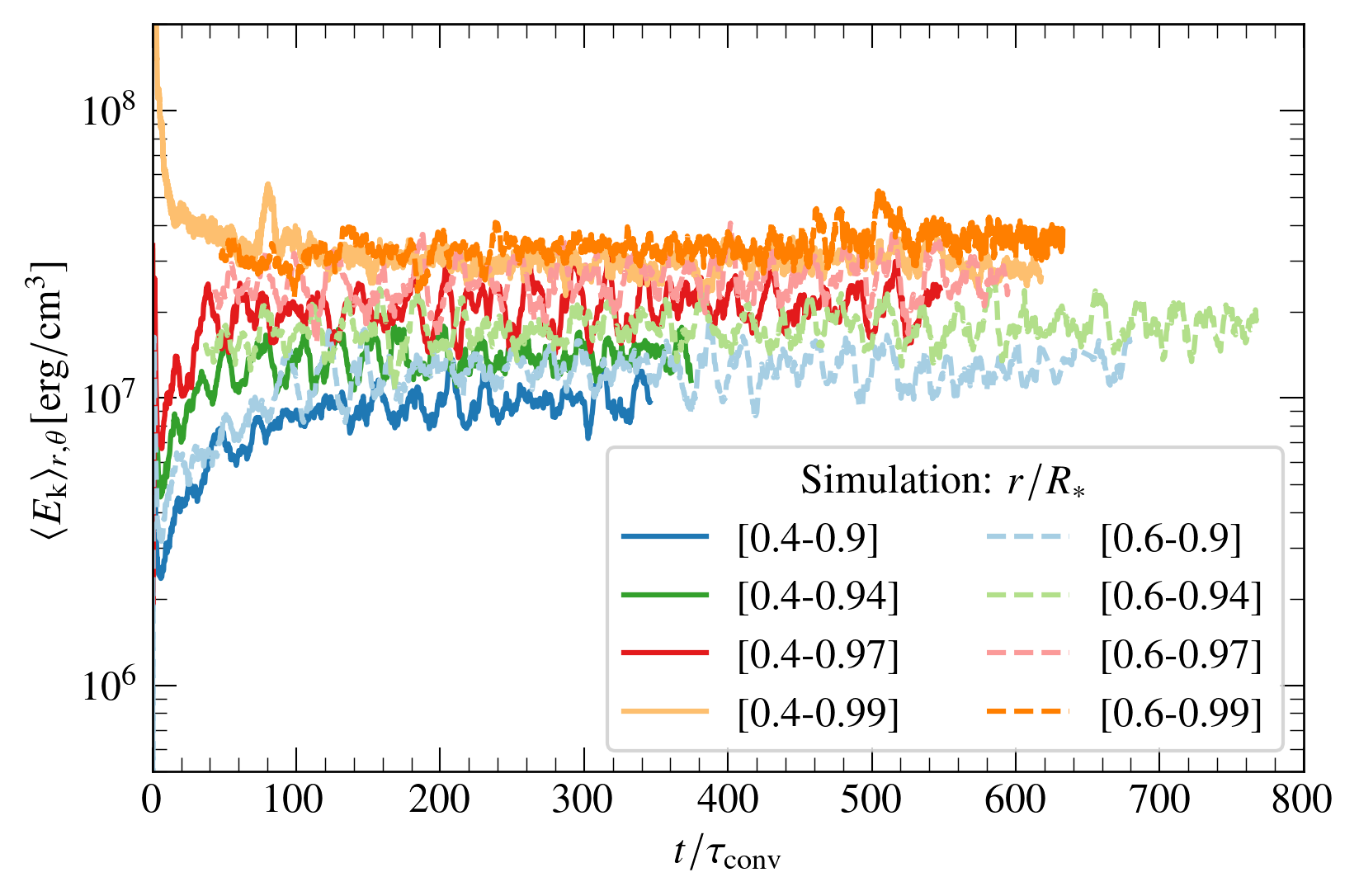}
	\caption{
	 Time evolution of the average kinetic energy density for all considered simulations.
	 Solid and dashed lines correspond to simulations with $\rmin = 0.4\rstar$ and $\rmax = 0.6\rstar$, respectively.
	 \label{fig:kinen_evol}
	 }
\end{figure}

The duration of each simulation is given in \cref{tab:params}.
As we will see in \cref{sec:results cz}, the typical velocity and hence global time-scale in the convection zone depends on the radial truncation.
So, in order to perform meaningful comparisons, we measure the evolution of the simulation in terms of the convective turnover time $\tconv$, which we define as
\begin{align}
\tconv = \int_{\rschw}^{\rmax} \frac{\di r}{ \avg{\vrms}_t},
\label{eq:tconv}
\end{align}
where $\vrms(r,t) = \sqrt {\avg{\vu^2(r, \theta, t)}_{\theta}} $ and $\avg{}_t$  and $\avg{}_\theta$ denote the time and volume-weighted angular average, respectively.
This estimates the typical time to cross the convection zone at the r.m.s. velocity.
As shown in \cref{fig:kinen_evol}, the evolution is characterised by an initial relaxation phase (approx. $100\, \tconv$) during which the spherically symmetric one-dimensional initial conditions relax and convection develops.
During the subsequent evolution the convection is quasi-steady
\eg in terms of the mean kinetic energy density $\avg{\ekin}_{r,\theta} = \avg{\rho \vu^2/2}_{r, \theta}$.
The value of $\tconv$ and all following results are based on the statistics of this quasi-steady period of evolution.

We collect statistics for a few hundred $\tconv$.
This is needed to ensure that the statistics in the convection zone are converged, since the overshooting layer is characterised by intermittent statistics \citep{brummell2002,pratt2017,baraffe2021}.
Thus, the long simulation time allows us to determine that the simulation is in a statistically stationary steady state.

Achieving exact thermal relaxation is a common challenge for global hydrodynamic simulations of convection based on realistic stellar interior structures \citep{meakin2007,horst2020,higl2021}.
The global thermal relaxation or Kelvin-Helmholtz timescale of the initial stellar structure used for our simulations is given by $\tau_{\mathrm{th}} = G M^2 /(R L)$, and ranges between \SI{5e4}{yr} and \SI{2e6}{yr}.
This is computationally unreachable without changing the underlying thermal time-scale, \eg by boosting the luminosity to shorten the KH time-scale.
However, as demonstrated by \citet{baraffe2021,lesaux2021} such boosting causes significant changes in the convection dynamics, the overshooting layer and the internal gravity waves in the radiative zone.
Consequently, we refrain from luminosity boosting in this study.
While the simulations are not completely thermally relaxed,
the evolution is very slow and smooth over the time period we investigate.
Thus, the time-scales of the processes we follow are significantly shorter than the simulation duration.

\section{Results: convection zone} \label{sec:results cz}

It has been proposed \citep{spruit1997,brandenburg2016,kapyla2017,anders2019} that stellar convection is primarily driven by the `rain' of radiatively cooled (low entropy) perturbations which originate at the top convective layers and buoyantly fall towards the lower convective boundary.
The perturbations traverse the bulk of the convection zone quasi-adiabatically.
It takes the boundary layer dynamics near the bottom convective boundary to break them up and extract their entropy deficit.
The flows through the convective zone of our simulation are qualitatively consistent with this picture, specifically in regards to the adiabatic transport through the convective bulk.

\begin{figure}
	\includegraphics[width=\columnwidth]{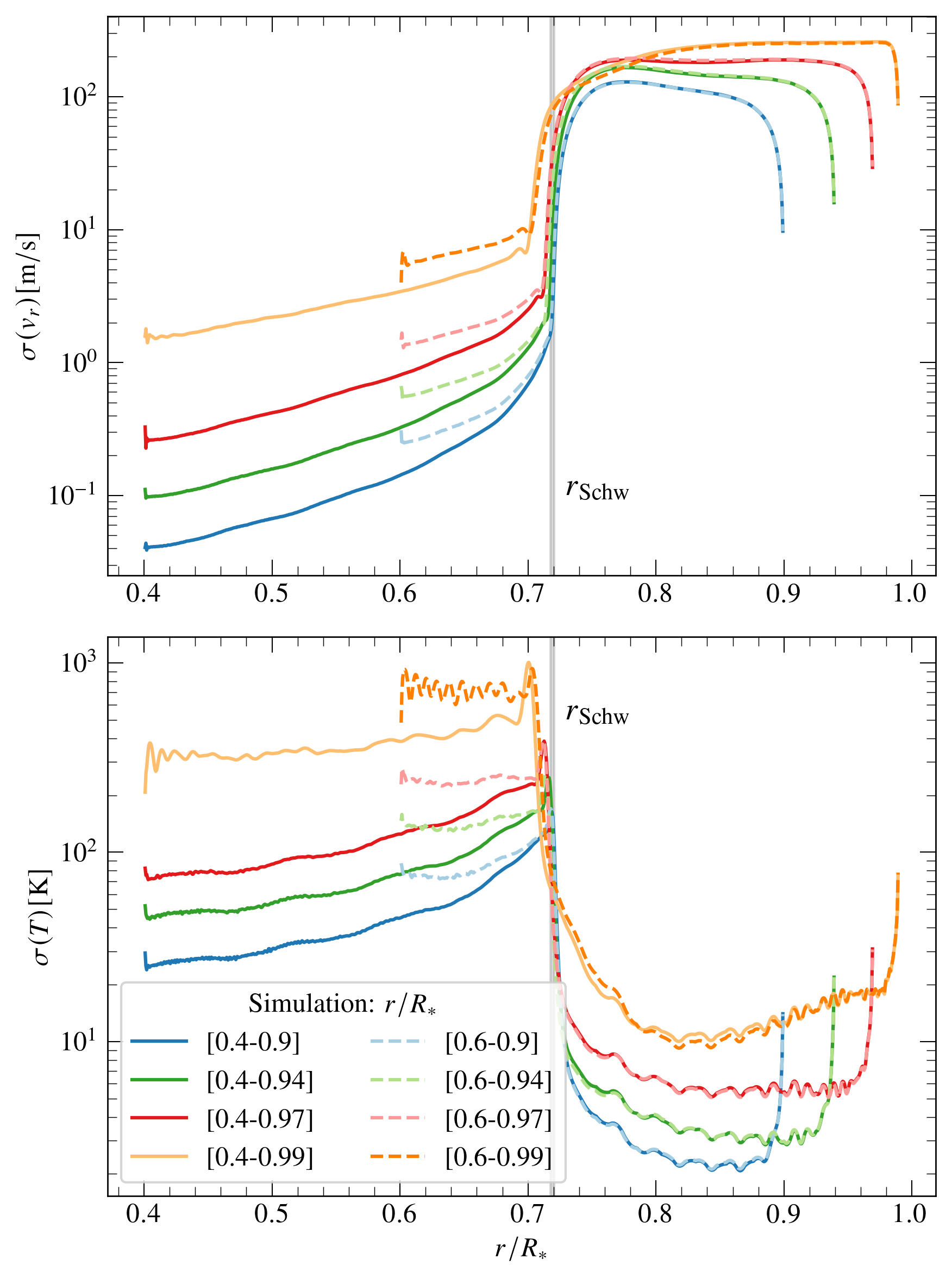} \hfill
	\caption{
	Radial profiles of  the standard deviation $\sigma$ of the radial velocity (top) and horizontal temperature fluctuations (bottom).
	The Schwarzschild boundary is marked by the vertical grey line.
	\label{fig:kinematics_cz}
	}
\end{figure}

\Cref{fig:kinematics_cz} shows the radial profiles of the standard deviation of the radial velocity  $\sigma(\vur ) = \sqrt{\avg{(\vur - \avg{\vur}_\theta)^2}_{t,\theta}}$ and the temperature fluctuations $\sigma(T) = \sqrt{\avg{(T - \avg{T}_\theta)^2}_{t,\theta}}$.
As expected, both profiles are almost constant in the bulk of the convection zone.
Because the background is stratified quasi-adiabatically, the latter indicates that the fluctuations also evolve quasi-adiabatically in the convective bulk.
This implies that the fluctuations generated near the boundaries do not have time to thermalise with the background as they travel through the bulk.
The ILES nature of the simulations prevents an exact quantification of the time scale of this thermalisation because of the numerical diffusion contribution. However, an upper bound is given by the thermal diffusion time scale $\tau_{\chi}$.
Here, we compute it based on the pressure scale-height as $\tau_\chi = H_p^2/ \krad$, with thermal diffusivity $\krad = \chi / (\rho c_p)$ and specific heat at constant pressure $c_p$.
Using this definition the thermal diffusion time scale is of the order of $\SI{2e12}{} - \SI{1e13}{s}$ in the convection zone increasing with increasing radius, \ie $7$ to $8$ orders magnitude larger than $\tau_\conv$ (see \cref{tab:params}).
Note that this is also consistent with previous results  \citep{viallet2015,pratt2016}.

The data show that increasing $\rmax$ increases both $\sigma(\vur )$ and $\sigma(T)$ (note the logarithmic axis in \cref{fig:kinematics_cz}).
At the same time shifting $\rmin$ by $20\%$ of the stellar radius (from $0.4\rstar$ to $0.6\rstar$) has an almost imperceptible effect on both diagnostics.
The same trend is observed for the average kinetic energy density (see \cref{fig:kinen_evol}) which is dominated by the convection zone.
This is consistent with the results of \citet{hurlburt1984} who studied a convection layer under similar boundary conditions but with an idealised setup (polytropic stratification, ideal gas EoS, Cartesian geometry, and maximum density contrast of $21$).
 They reported a growth of the mean kinetic energy with the number of pressure scale heights in the simulation domain.

At first glance the dependence on $\rmax$ and the insensitivity to $\rmin$ that we report may be surprising because all simulations are driven with the same luminosity and share the same stratification and boundary conditions.
Indeed, the main energy balance in steady state in the convection zone is between the kinetic energy and convective heat fluxes, which must add up to transport the total stellar luminosity.
However, due to mass conservation and the density stratification, the mean kinetic energy flux is negative.
Consequently, the energy balance is not sufficient to determine the amplitudes of the two fluxes.
Thus, the amplitudes of the velocity and temperature fluctuations are allowed to vary with $\rmax$ while maintaining the same luminosity.
The dependence
can be explained by the properties of the background at the top convective boundary, which change with $\rmax$.
 For instance the background temperature drops from $T(\rmax=0.9\rstar) =  \SI{6.1e5}{K}$ to
 $T(\rmax = 0.99\rstar) = \SI{0.51e5}{K}$,
and the density drops from $\rho(\rmax=0.9 \rstar) = \SI{23.7e-3 }{g .cm^{-3}}$ to $\rho(\rmax = 0.99\rstar) = \SI{0.3e-3}{g. cm^{-3}}$, see \cref{tab:params}.
 At the same time, the properties of the background at the lower convective boundary are largely the same across all simulations (\eg $T(\rschw) = \SI{2.15e6}{K}$ and $\rho(\rschw) = \SI{0.16}{g.cm^{-3}}$).
This implies that for a fixed realistic stellar background, the properties of the radial velocity and temperature fluctuations are determined by the conditions in the boundary layers, at least for the type of convection zone that the simulations can represent.
This leads us to consider the outer boundary layer in more detail.
Specifically, we consider the fluctuations at $\rmax$ and how they evolve across the boundary layers.

As we evolve the specific internal energy density as a primitive variable, $\sigma (T) (\rmax)$ is implicitly set by the stratification and the imposed luminosity.
The larger value of $\sigma (T) (\rmax)$ for increasing $\rmax$ can be primarily attributed to the corresponding decrease of the mean heat capacity $\avg{\rho c_p}_{t, \theta}$.
The luminosity is carried predominantly\footnotemark{} by the convective heat flux in this region.
Hence, as the mean heat capacity at $\rmax$ decreases across simulations, $\sigma (T) (\rmax)$ must increase to maintain the same luminosity.

\footnotetext{Because the radial velocity and the temperature are computed on staggered grids in \MUSIC, the radial velocity and hence the convective heat flux interpolated to the centre of the outermost $\rmax$ cell can be non-zero. At this location, the radiative diffusivity is negligible and hence the radiative flux is insignificant.}

To describe the radial structure of the temperature and radial velocity fluctuations at the outer boundary, we use the linear rescaling
 \begin{align}
\hat{\sigma} = \frac{\sigma(r) - \sigma_{\bulk}}{|\sigma(\rmax) - \sigma_{\bulk}|},
\label{eq:bl_rescale}
 \end{align}
where $\sigma$ stands in for $\sigma(T)$ and $\sigma(\vur)$ and
 $\sigma_{\bulk}$ is the volume-weighted mean value of the respective $\sigma(r)$ in the convective bulk.
Here, we consider the convective `bulk' to consist of the shell $[\rmax - 0.08\rstar, \rmax - 0.02\rstar]$.

\begin{figure}
	\includegraphics[width=\columnwidth]{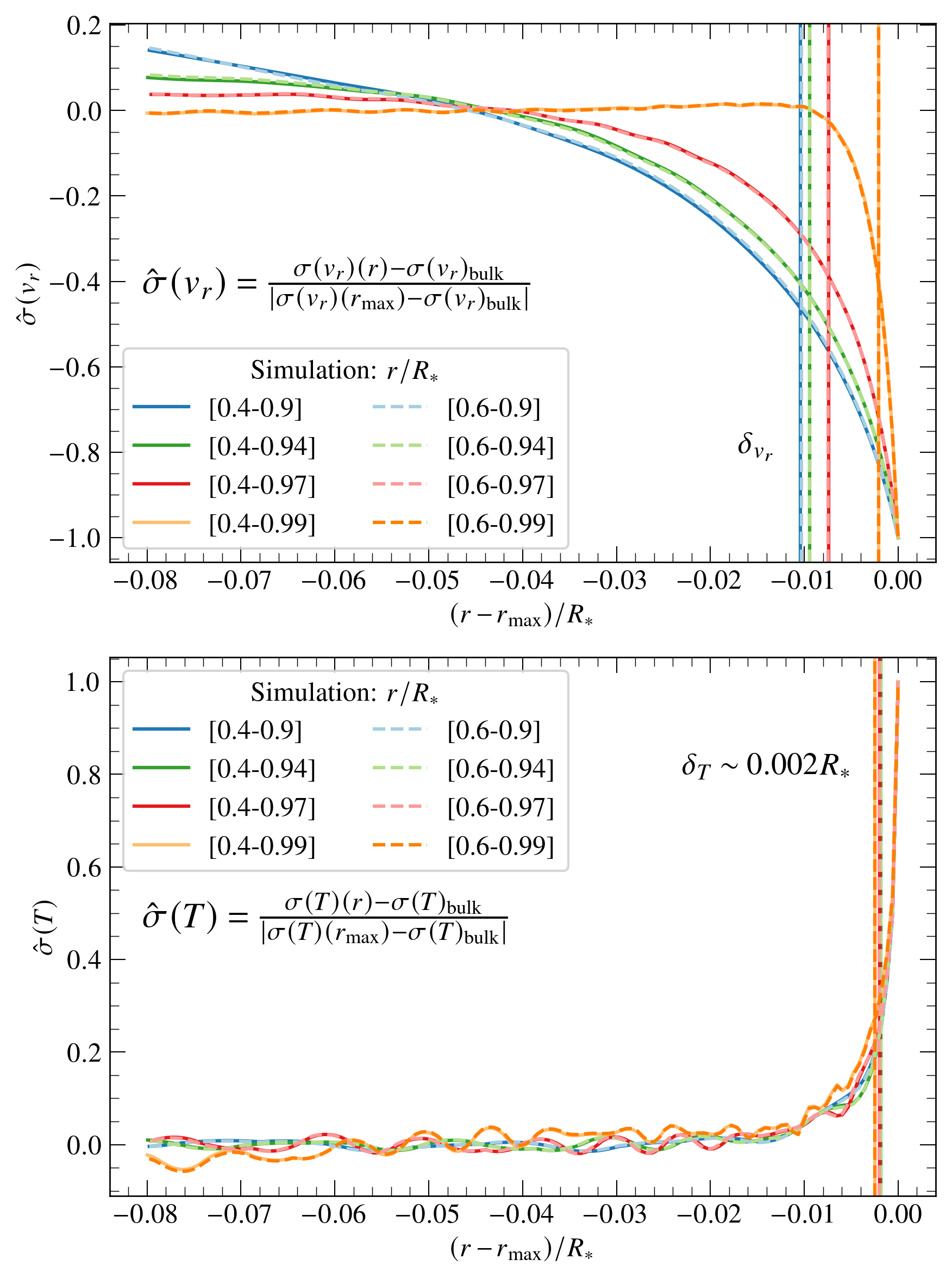}\\
	\caption{
	Radial profiles of temperature fluctuations (top) and radial velocity (bottom) in the upper boundary layer linearly rescaled as indicated by \cref{eq:bl_thickness}. The vertical lines indicate the corresponding boundary layer thickness, see \cref{eq:bl_thickness}.
	\label{fig:surf_BL}
	}
\end{figure}

\Cref{fig:surf_BL} shows $\hat{\sigma}(T)$ and $\hat{\sigma}(\vur)$ along with the associated boundary layer thickness
defined similarly to \citet{Featherstone2016}  as
\begin{align}
\delta = \int_{\rmax - 0.08 \rstar}^{\rmax} \hat{\sigma}(r) dr.
\label{eq:bl_thickness}
\end{align}
Here we set the lower integration boundary well above the lower convective boundary, so that $\delta$ is not influenced by the boundary layer near $\rschw$.
As the figure shows, the hydrodynamic boundary layer becomes significantly thinner with increasing $\rmax$, as indicated by $\delta_{\vur}$, see also \cref{tab:params}.
In contrast, the rescaled thermal fluctuations $\hat{\sigma}(T)$ have the same radial dependence across all simulations with a thermal boundary layer thickness $\delta_T = 0.002 \rstar$.
As a result, the ratio $\delta_{\vur}/\delta_T$  decreases with increasing $\rmax$ reaching unity for the  $\rmax = 0.99\rstar$ simulations.
This can explain the $\rmax$ dependence of  $\sigma(\vur)$ and $\sigma(T)$ in the convective bulk, as follows.

By definition, the hydrodynamic boundary layer is characterised by the growth of the radial velocity from zero to its bulk value, due to \eg buoyancy.
Similarly, the thermal boundary layer is characterised by the decay of temperature fluctuations through \eg diffusion and horizontal shear.
Thus, as $\delta_{\vur}/\delta_T$ increases the hydrodynamic boundary layer is dominated by smaller temperature (and hence density) perturbations and thus weaker buoyancy.
At the same time, the thermal boundary layer is characterised by weaker radial advection and hence a longer time scale over which the temperature fluctuations can decay.
Hence, as $\delta_{\vur}/\delta_T$ increases the saturation values of both the temperature and velocity fluctuations, \ie their values in the convective bulk,  drop.
The precise values of $\delta_{\vur}$ are $\delta_T$ are a result of a detailed balance between radial advection and buoyancy, on one hand, and diffusion, horizontal shear and dissipation on the other.
As is typical for global stellar hydrodynamic simulations, the ILES nature of the simulations precludes an in-depth discussion of this balance.
However, we note that
the boundary layers thicknesses are not set by the local stratification, \eg they do no scale with $H_p(\rmax)$.
Moreover, the measured value of $\delta_T$ is limited by the grid resolution in all simulations (corresponding to approximately 3 grid cells), while $\delta_{\vur}$ is only resolution-limited in the largest $\rmax$ simulations.
Thus, the ratio $\delta_{\vur}/\delta_T$ should increase with increasing radial resolution.
We establish this by repeating the $0.4\rstar - 0.94\rstar$ and $0.6\rstar-0.99\rstar$ simulations with $\sim \SI{50}{\percent}$  larger $N_r$, while keeping all other parameters the same.
The simulations start from the same initial conditions and evolve for $\SI{1.8e8}{s}$.
The thermal boundary layer shrinks in physical units with the increased resolution, so that $\delta_T$ remains approximately the same number of radial grid cells.
However, the $\delta_{\vur}$ remains approximately the same thickness (in physical units) for the $0.4\rstar - 0.94\rstar$ truncation
and it grows by approximately $\SI{20}{\percent}$ for the $0.6\rstar-0.99\rstar$ truncation.
Importantly, the radial profiles of $\sigma(T)$ and $\sigma(\vur)$ in the convective bulk are not affected by the small increase in resolution.

In  summary, the thermal perturbations originating in the upper boundary layer are effectively frozen while they travel through the convective bulk, because of the long diffusion time scale.
As a result the speed and temperature contrast of the radial flows near the lower boundary layer is effectively set by the conditions at the upper boundary layers and not in the convective bulk.
The key parameters are the density stratification and the ratio between the hydrodynamic and thermal boundary layer thicknesses.
For this set of simulations, the density stratification is steep enough to lead to decreasing heat capacity with height, which requires larger temperature fluctuations to maintain the same luminosity.
In addition, the ratio of  $\delta_{\vur}/\delta_T$ decreases towards unity with increasing $\rmax$,
meaning that an increasingly bigger part of the radial velocity is generated by not-fully-decayed thermal fluctuations with larger buoyancy.
This leads to $\sigma(T)_{\bulk}$ which scales with $\sigma(T)(\rmax)$ and $\sigma(\vur)_{\bulk}$ which increases correspondingly.
This qualitative picture illustrates how the boundary layer dynamics can determine the properties of the convective bulk and illustrates the non-locality of boundary-driven convection.
It is also consistent with the results of \citet{cossette2016}, who find that for boundary-driven convection the initial density and entropy contrast of the adiabatically descending perturbations determine the dynamical properties of the convective bulk.

\section{Overshooting layer} \label{sec:results os}
\subsection{Characteristic overshooting depth} \label{sec: overshoot depth}
As mentioned, the background properties near the lower convective boundary are the same among all simulations.
However,  as we show in this section, the non-locality of convection means that the $\rmax$-dependence of the outer convective layers makes a qualitative difference to the
lower convective boundary layer dynamics and convective overshooting in particular.

We recall that we use the Schwarzschild criterion to locate the lower convection boundary.
The overshooting layer is then the region below this boundary where convection flows are able to penetrate.
Specifically, we characterise the overshooting in the same way as \cite{baraffe2021}.
Namely, we consider the radial kinetic energy $F^{\ke}_r$ and convective heat flux $F^{\conv}_r$, defined by
\begin{alignat}{3}
	\v{F}^{\conv}_r(r, \theta, t)& = h' (\rho \vur)',
	\label{eq:flux_h ave}\\
	\v{F}^{\ke}_r(r, \theta, t)& = \vur \ekin = \vur \half \rho \vu^2,
	\label{eq:flux ave}
\end{alignat}
where $h= e + p/\rho$  is the specific enthalpy and the primes denote fluctuations from the angular average, \eg $h' = h - \avg{h}_\theta$.
Here, the convective heat flux contains only the enthalpy fluctuations.
From the hydrostatic energy balance, \ie hydrostatic equilibrium, it can be seen that the mean enthalpy balances the gravitational potential
(see \cref{app:h_flux}).
We consider $F^{\conv}$ because our simulations solve the internal energy equations where the enthalpy appears naturally.
A quantity that is also often used in the literature is $\avg{F_{\Delta T}}_{t, \theta} = \avg{\rho c_p \Delta T \vur }_{t, \theta}$ \citep{hurlburt1984,zahn1991,baraffe2021}, where $\Delta T$ is the temperature fluctuations with respect to a reference model.
It is closely related to $\avg{F^{\conv}_r}_{t, \theta}$ (as can be seen analytically e.g. for an ideal gas), and in the considered simulations has the same quantitative behaviour.

To identify an instantaneous local overshooting event, we track the locations of the first zero-crossing of the two fluxes $F^{\conv}_r$ and $F^{\ke}_r$ below the Schwarzschild boundary, $l^{\ke}(\theta, t)$ and $l^{\conv}(\theta, t)$ respectively.
Specifically, we consider the instantaneous position of the boundary $\rschw(t) =  \min \{r: \avg{\nabla - \nablaad}_\theta  \geq 0 \}$.
Retaining the time-dependence of $\rschw(t)$ here allows us to extract
the effects of the convection motion in the stably-stratified region and exclude the motion of the convective boundary itself.
\Citet{brummell2002} distinguishes the two as `convective overshooting' and `convective penetration', respectively.
Note, however that the time variation of the boundary $\rschw(t)$ is very slow (significantly slower than the $\tau_{\conv}$ and much slower than the time scale of individual overshooting flows) and as such, it would be perceived as stationary by individual penetrating flows. We discuss the modification of the background and possible convective penetration in more detail in \cref{sec:os_bkgr}.

The distribution of both $l^{\ke}$ and $l^{\conv}$ can be associated with significant frequency of extreme events both in time and space \citep{pratt2017}.
We confirm this in the presented datasets,
so we use two summary statistics to describe the overshooting
\begin{align}
l_{\bulk}(t) &= \avg{l(\theta, t)}_{\theta}, \\
l_{\max}(t) &= \max_{\theta} \{ l(\theta, t) \}.
\end{align}
Here $l_{\bulk}$ describes the angular average and is interpreted as the \emph{typical} depth of overshooting,
while $l_{\max}$ is the angular maximum, \ie the instantaneous maximum depth of overshooting
and $l(\theta, t)$ can refer to either $l^{\conv}(\theta, t)$ or $l^{\ke}(\theta, t)$.
\begin{figure*}
		\includegraphics[width=\textwidth]{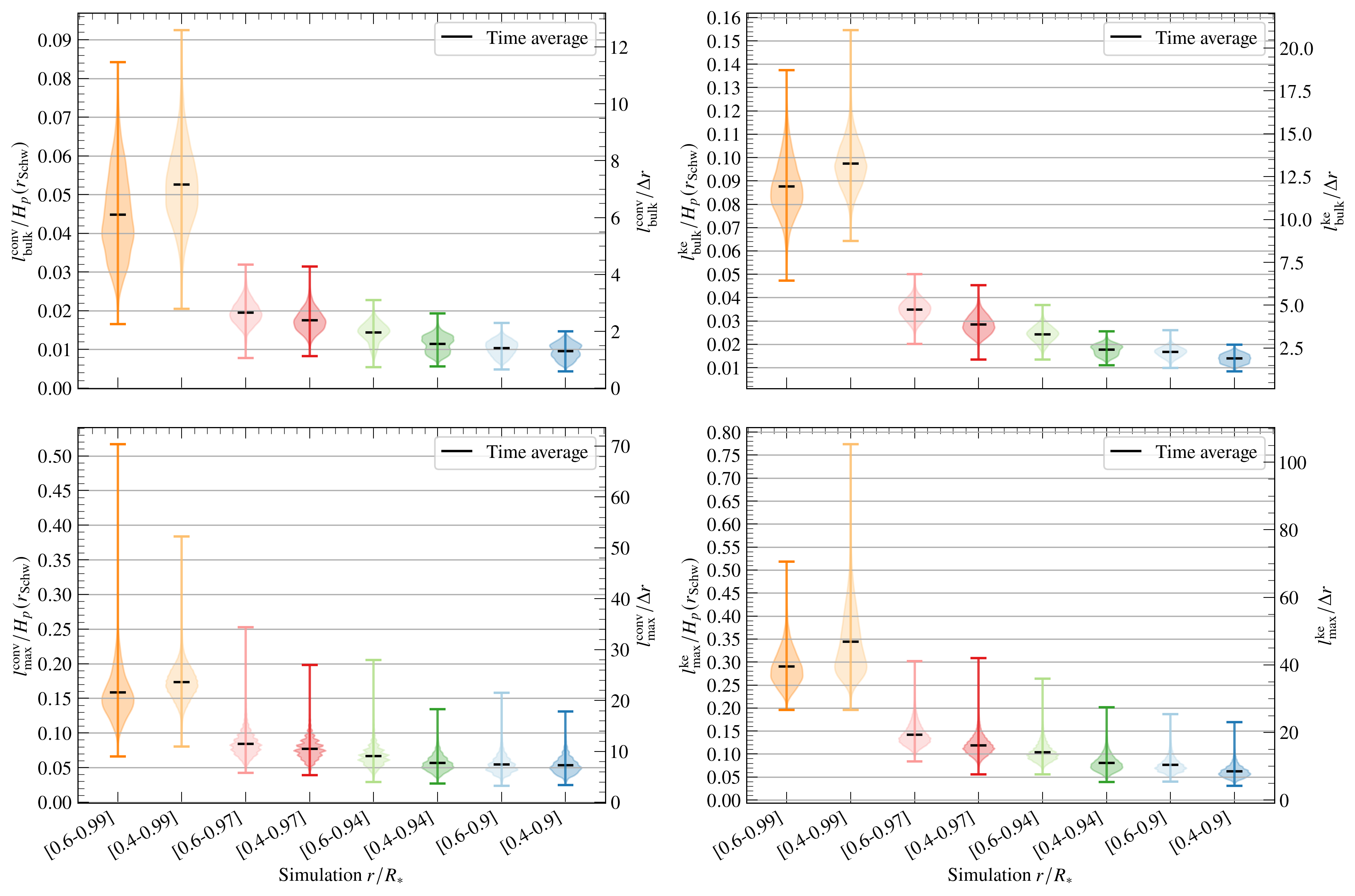}\\

	\caption{
	Time distribution of the instantaneous overshooting depth in units of the pressure scale height at the convective boundary (left axis) and radial cell width (right axis):
	mean-in-$\theta$ $l_{\bulk}$ (top row) and  max-in-$\theta$ $l_{\max}$ (bottom row).
	The left hand side panels use the convective heat flux and the right column -- the kinetic energy flux, \ie the radial velocity.
	The simulations are ordered by decreasing $\rmax$ and $\rmin$.
	}
	\label{fig:d_os_t}
\end{figure*}
\Cref{fig:d_os_t} shows the distribution of these quantities across time in the different simulations in units of the pressure scale height at the Schwarzschild boundary $H_p(\rschw)$.
We exclude a $\ang{15}$ angle around the polar axes, to reduce the impact of the polar boundaries
-- the reflective boundary conditions drive strong radial flows which result in locally deeper overshooting.
The figure shows that there is a clear dependence of the overshooting depth on $\rmax$ -- larger $\rmax$ is associated with a thicker overshooting layer.
There is also such a dependence for both $l_{\bulk}$ and $l_{\max}$,  based on either of the fluxes, independent of whether we consider the time-average or either of the tails of the temporal distribution.
At the same time there does not seem to be a clear dependence on $\rmin$ shared by all overshooting depth diagnostics, for the two $\rmin$ we examine.

These observations correlate well with the dependence of the convective intensity (as measured by \eg $\sigma(\vur)$ and $\sigma(T)$, \ie  the radial velocity and temperature fluctuations) on $\rmax$ and $\rmin$, see \cref{fig:kinematics_cz}.
This suggests that the larger overshooting depth is caused by more vigorous convection, \ie faster and colder downflows can reach deeper into the stably-stratified region and such flows are generated in simulations with larger $\rmax$.
For instance, in the $\rmin = 0.4\rstar$ simulations $\sigma(\vur) (\rschw)$ takes the values $[\SI{5e2}{}, \SI{1e3}{}, \SI{3e3}{}, \SI{8e3}{}]$ $\si{cm. s^{-1}}$
for $\rmax/\rstar = [0.9, 0.94, 0.97, 0.99]$,  respectively.
The lack of dependence on $\rmin$ indicates that either the feedback from the radiative zone is a negligible influence on the overshooting, or that the feedback itself does not depend very strongly on the spectral properties of internal gravity waves, at least up to the variations of the IGWs across the presented simulations.
This will be discussed in more detail in \cref{sec:results rad}, where we show that the spectrum of the travelling waves does not depend qualitatively on $\rmin$.

It follows trivially from the definitions \Cref{eq:flux_h ave,eq:flux ave} that the loci of $l^{\conv}(\theta, t)$ and $l^{\ke}(\theta, t)$ can be identified with the first zero-crossings of the vertical momentum and enthalpy fluctuations.
In other words, they track the bottom end of the downflows as defined by their velocity and thermal properties.
Furthermore, by construction $l^{\ke} \geq l^{\conv}$, because $l^{\ke}$ depends only on the zeroes of the radial velocity, while $l^{\conv}$ is also triggered by the zeroes of the enthalpy fluctuations.
Physically, this is because  the radial buoyant acceleration vanishes at $l^\conv(\theta, t)$, while the radial velocity vanishes at $l^{\ke}(\theta, t)$.
The inequality is not saturated in the presented simulations despite the long run times.
Even instantaneously, the difference between the bulk values $l^{\ke}_{\bulk}(t) - l^{\conv}_{\bulk}(t)$ remains steady at about $\SI{8}{\percent}$ to $\SI{10}{\percent}$ of their mean value $(l^{\ke}_{\bulk}(t) + l^{\conv}_{\bulk}(t))/2$.
The difference between the extreme values $l^{\ke}_{\max}(t) - l^{\conv}_{\max}(t)$ is approximately  $\SI{20}{\percent}$ to \SI{30}{\percent} of the corresponding mean.

This indicates that typically the temperature fluctuations dissipate before the downflows stop descending (whereby we consider the statistical variation to be negligible).
While buoyancy is sufficient to explain this, it is not the only the process acting on the overshooting flows.
More generally for the inequality $l^{\ke} > l^{\conv}$ to be strict, the mechanisms removing the heat deficit of downflows must operate on shorter times-scales than those dissipating the excess radial momentum.
Broadly speaking, the heat deficit can be removed by adiabatic compression and heat exchange with the background and the IGWs, \eg enthalpy mixing/entrainment and radiative flux, wave excitation and breaking.
The radial momentum excess is removed by buoyancy, turbulent mixing, viscous dissipation, and IGW excitation.

\subsection{Modification of the thermal background} \label{sec:os_bkgr}
\begin{figure}
		\includegraphics[width=\columnwidth]{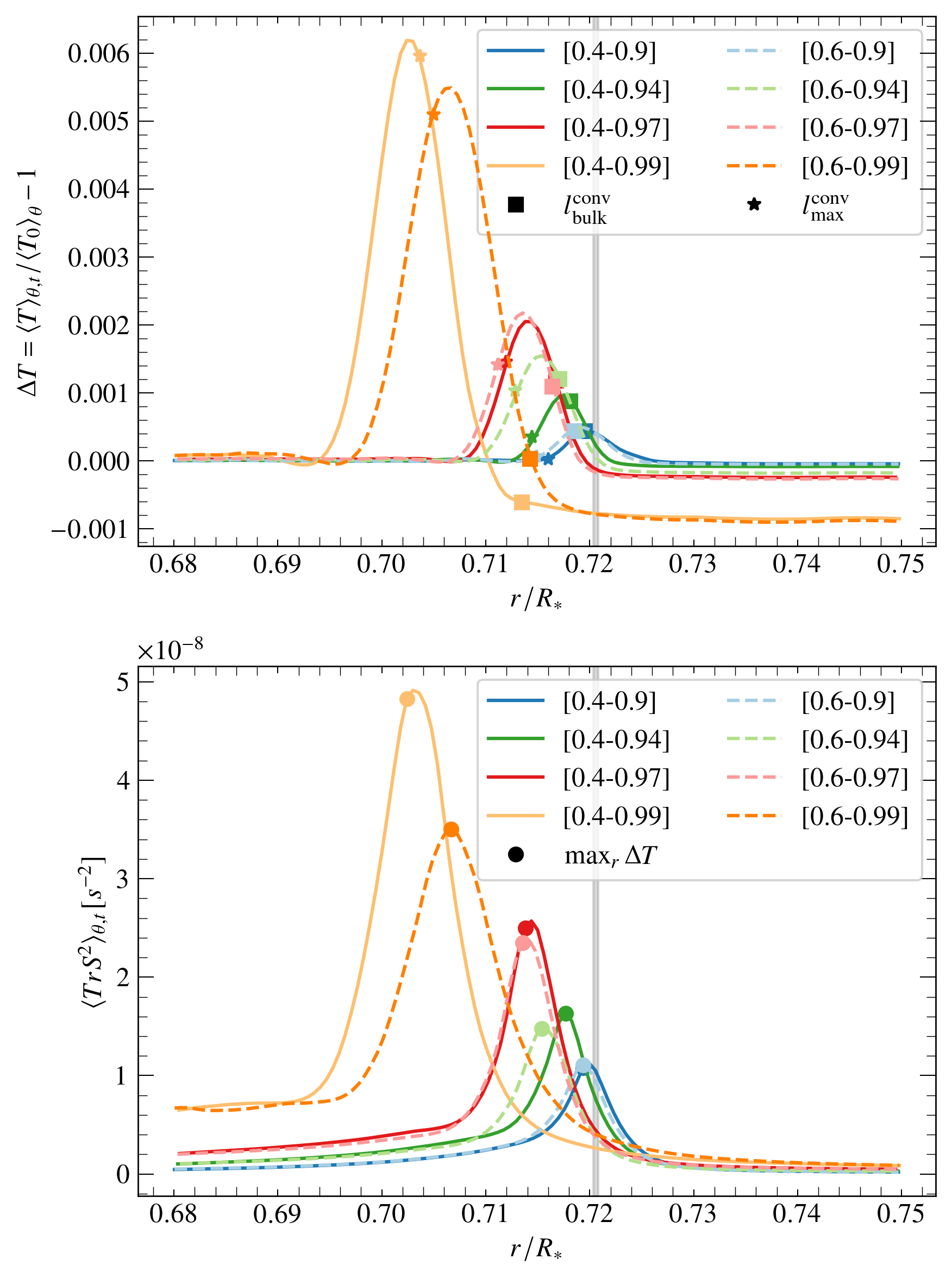}\\

	\caption{
	(top) Relative difference $\Delta T$ in the mean temperature background in the quasi-steady state $\avg{T}_{\theta, t}$ from the background at the beginning of the quasi-steady state $\avg{T_0}_{\theta}$.
	The locations of the overshooting depth measured by the convective heat flux are marked for reference.
	(bottom) Radial profile of the squared rate-of-strain averaged over $\theta$ and time. The location of the peak $\Delta T$ is marked with  $\bullet$ as a visual guide.
	The vertical grey line in both panels corresponds to the Schwarzschild boundary.
	}
	\label{fig: deltaT}
\end{figure}

As the downflows travel below their local, instantaneous $\rschw(\theta, t)$ depth, and the background stratification becomes sub-adiabatic, they become hotter than the background.
Some of this excess heat may be deposited in the overshooting layer through irreversible processes (e.g. IGW excitation/breaking, mixing, etc.).
Indeed, as \cref	{fig: deltaT} shows, there is a good correlation between the overshooting depth and a layer which is slightly hotter than the
background at the beginning of the quasi-steady state.

Indications for a similar process (a layer with a negative enthalpy flux, an excess radiative flux, or slightly larger mean temperature) has been previously reported by \eg \citet{rogers2006,brun2011,brun2017,korre2019,baraffe2021} and \citet{higl2021} in a variety of parameter regimes (see \citealt[section 3.2]{baraffe2021} for more details).

\Citet{baraffe2021} note that there is a good correlation between the location of the radial profile of the heating layer and the trace of the square rate-of-strain $\Tr S^2 = \Tr (\nabla \vu + \nabla \vu^T)^2/4$, where $\nabla \vu$ denotes the velocity gradient.
We can confirm this correlation as well, see \cref{fig: deltaT}.
Note that $ \Tr S^2$ encodes two types of material deformation -- dilatation and shear.
However, because of the low Mach number $\la 10^{-3}$, $\Tr S^2$ is dominated by the incompressible shear component (not shown).
This suggests that the primary component of the heat deposition is shear-induced mixing of previously adiabatically compressed hot material.

This local heating causes a dip in the sub-adiabaticity $\nabla - \nablaad$ just below the convective boundary and a corresponding increase below the dip, around the heat bump.
If left unstopped this process may lead to the so-called convective penetration in the language of \citet{brummell2002}, \ie the descent of the convective boundary and growth of the convection zone.

Energetically, in a sub-adiabatically stratified medium adiabatic compression of downflows and adiabatic expansion of upflows both lead to downward/negative enthalpy flux
\citep{hurlburt1994,muthsam1995,brun2011,pratt2017,korre2019,kapyla2019}.
So, as discussed in \citet{baraffe2021}, the reason for the local change in the background stratification
must be that the heat deposited by the enthalpy flux cannot be efficiently evacuated.
Considering the internal energy equation, \cref{eq:eint},  the excess heat deposited by the enthalpy flux can only be evacuated through thermal diffusion, \ie radiative flux (considering that the local Mach number is $\la 10^{-3}$ the $p \div \vu $ term does not play a significant role).
Indeed, as noted by \citet{baraffe2021}, the local heating must lead to increased thermal diffusivity and hence increased radiative flux.
The feedback should eventually grow sufficiently to balance the negative enthalpy flux.
Indeed, we note a slow growth in the heating rate throughout the simulated time interval\footnotemark.
However, such a regime cannot be reached with the presented simulations.

\footnotetext{
The available data is consistent with a constant convective heat flux and a linearly growing radiative flux in the overshooting layer. Extrapolating this in a zero-order approximation,
the two fluxes would match and the excess heating would saturate on a time scale of $O(\si{10^9 s})$.
}

\section {Radiative zone} \label{sec:results rad}

One of the effects of the convection zone and the overshooting layer is to excite internal gravity waves (IGW) in the radiative zone.
Internal gravity waves play a crucial role in the internal structure and evolution of stars,
as they redistribute angular momentum, energy and chemical species,  deposited in the overshooting layer by convective flows.
It is interesting then to consider how the changes in the overshooting layer and the convection zone due to the different radial truncations are reflected in the IGWs, as well as the strength of any potential feedback onto the convective overshooting.

We focus the analysis on the power spectra of radial velocity.
The spectral coefficients $\widehat{\vur}(\ell, \omega)$ are obtained using the conventions in \citet{lesaux2021}.
In short, we perform a Blackman-windowed Fourier transform with respect to time and a 2D spherical harmonic transform ($m=0$) with respect to the angular direction $\theta$.
The resulting 2D power spectrum is then given by
\begin{align}
\mathcal{P}[\vur](\ell, \omega) = \begin{cases}
\abs{\widehat{\vur}(\ell,\omega)}^2     & \textrm{for } \omega = 0 \\
2\abs{\widehat{\vur} (\ell, \omega)}^2 & \textrm{for } \omega >0,
\end{cases}
\end{align}
where $\ell$ is the angular degree and $\omega$ is the frequency.

\begin{figure*}
	\includegraphics[width=\textwidth]{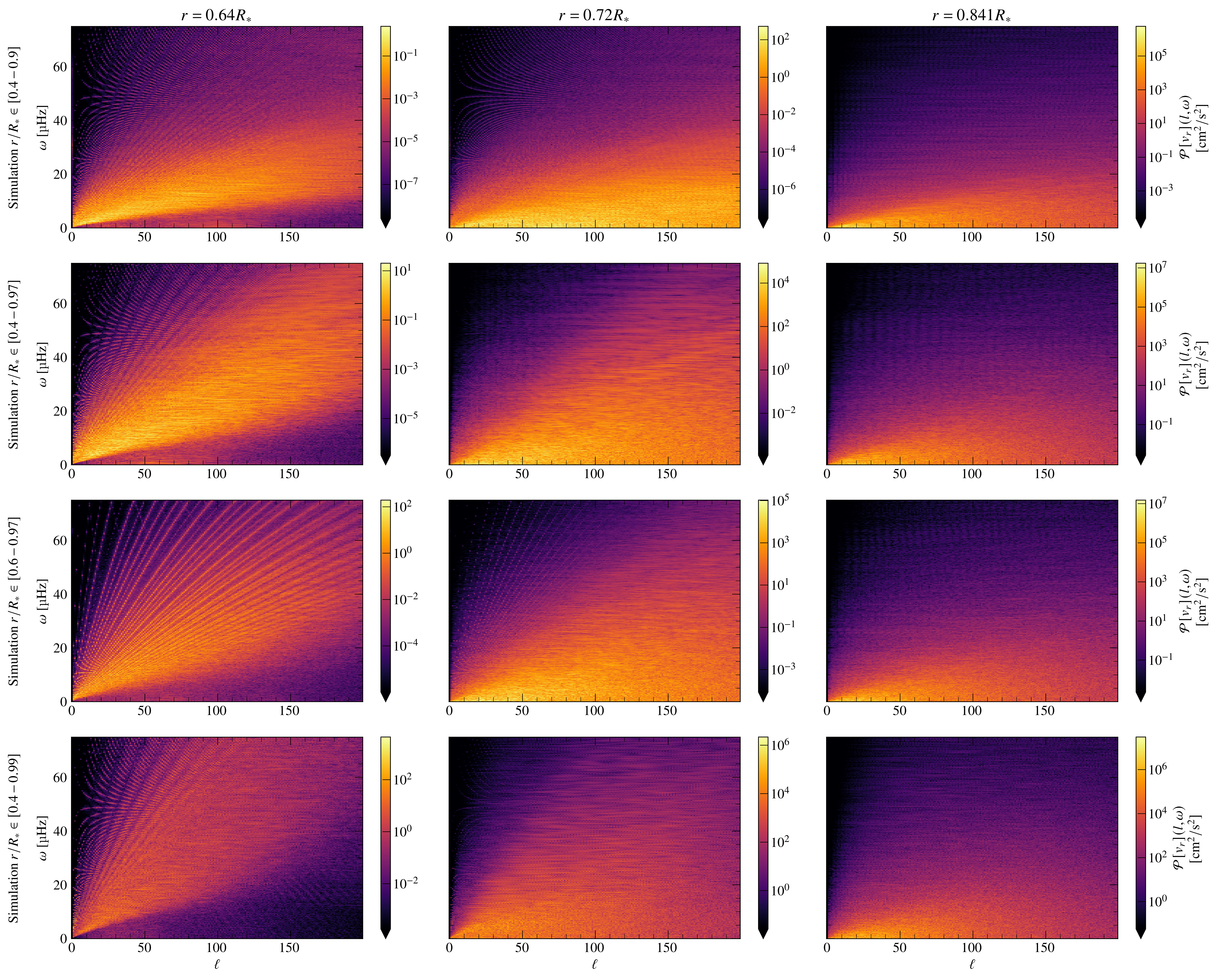}\\

	\caption{
	Power spectra of the radial velocity deep in the radiative zone -- $ H_p(\rschw)$ below the Schwarzschild boundary (left),
	at the bottom end of the mixing zone at $l_{\max}$ (middle), and in the bulk of the convection zone $0.841 \rstar$ (right).
	Each row corresponds to spectra from the same simulation.
	All spectra are produced from data spanning the last $\SI{2.25e7}{s} $ of each simulation.
	The lower five percentile of the data are rendered in black to improve the colour contrast and aid visual comparison.
	The frequency axis extends up to $3/4$ of the Nyquist frequency $\Ny$ to avoid aliasing artefacts. $\Ny = (2\delta t)^{-1} =  \SI{100}{ \micro Hz}$ based on separation between snapshots of $\delta t \sim \SI{5e3}{ s}$.
	Note that the power deficit at the low-$\omega$ high-$\ell$ region of the $r = 0.64 \rstar$ panels is associated with resolution constraints related to the finite duration of the analysed data and the cell size.
	}
	\label{fig:spec2d}
\end{figure*}

 \Cref{fig:spec2d} shows the radial velocity power spectra for a representative selection of simulations at three locations.
 The spectra are computed from a subset of the data spanning $\SI{2.25 e7}{s}$ in the quasi-steady state.
The deepest location is chosen in the bulk of the radiative zone at $r=0.64 \rstar = \rschw - H_p(\rschw)$  far away from any overshooting activity and we expect it to be dominated by the IGW signal.
The intermediate location is at the convective boundary at $\rschw$ and as such the spectra there are expected to exhibit signs of the overshooting dynamics superposed with the IGW signal.
The highest location is in the bulk of the convective zone at $r= 0.84\rstar$ and is shown as a reference to illustrate the length- and time-scales of the convective flow which is responsible for exciting the IGW signal.

The classical pattern of IGWs \citep{alvan2014,horst2020} is clearly visible in the radiative zone for all simulations, most prominently in the low-$\ell$ range.
This pattern can be inferred from the dispersion relation ${\omega^2}/{N^2} = {k_h^2}/{\v{k}^2}$ (see e.g. \citet{press1981}), where $\v{k}$ is the wave vector and $k_h^2 =\ell \bra{\ell +1}r^{-2}$ its horizontal component. After re-arranging, the dispersion relation can be seen as
\begin{align}
\bra{\bra{\frac{ 2 \ell+1}{2r k_r}}^2+ 1 - \frac{1}{4 r^2 k_r^2} }\bra{1 - \frac{\omega^2}{N^2}}  = 1,
\label{eq:disprel}
\end{align}
 which describes a family of hyperbolae in $(2\ell+1)^2$ and $\omega^2$ for a fixed $r$ (parametrised by $k_r$).

\begin{figure*}
	\includegraphics[width=\textwidth]{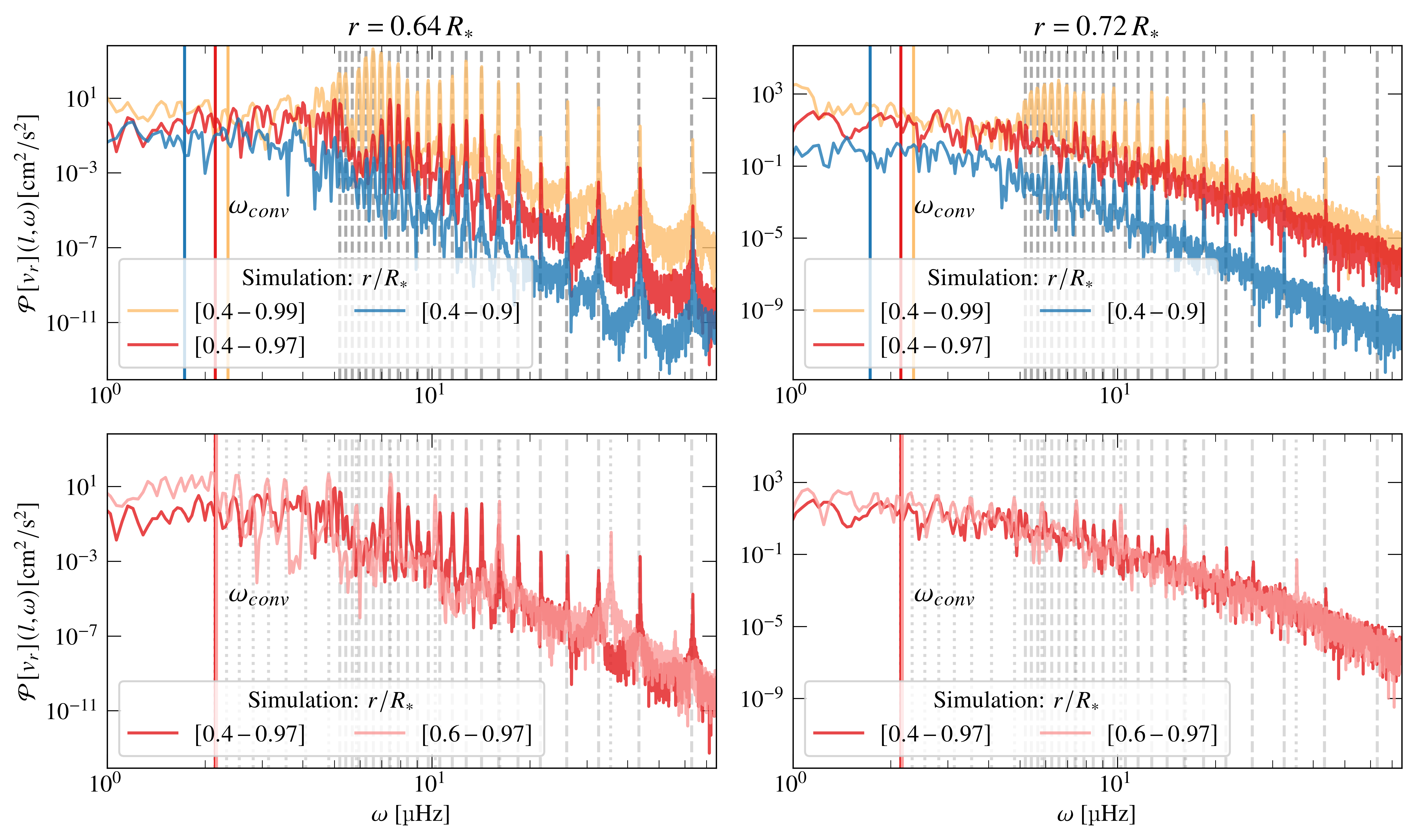}\\

	\caption{
	Frequency distribution of the radial velocity power spectrum $|\widehat{\vur}|^2 (l=2, \omega)$
	deep in the radiative zone (left) and at the convective boundary $\rschw$ (right).
	Top row shows a comparison of simulations with $\rmin = 0.4 \rstar$ and the bottom row -- with $\rmax = 0.97 \rstar$.
	The dashed and dotted vertical lines show the frequencies of the $g$-modes obtained with GYRE for the $\rmin=0.4\rstar$ and $\rmin=0.6\rstar$ respectively.
	The solid coloured vertical lines show the convective frequency $\omega_{\conv} = 1/ \tau_{\conv}$ for the respective simulation.
	}
	\label{fig:spec1d_l2_GYRE}
\end{figure*}

To confirm quantitatively that the computed spectra in the radiative zone contain IGWs,
we compute the frequencies of the $g$-modes (\ie the standing waves) associated with the radiative zones of the simulation for a few distinct values of $\ell$.
The computation is performed using the stellar oscillations code GYRE%
\footnote{Version 6.0, see \href{https://gyre.readthedocs.io}{https://gyre.readthedocs.io}}
\citep{townsend2013,townsend2018}.
Since the modes depend only on the stratification and the geometry of the resonant cavity (\ie the radiative zone), we obtain two distinct sets of modes for the $\rmin=0.4\rstar$ and $\rmin=0.6\rstar$ sets of simulations.
As \cref{fig:spec1d_l2_GYRE} shows for the $\ell=2$ modes, the frequencies obtained with GYRE match almost perfectly the peaks in the spectra for both values of $\rmin$ in the radiative zone and at the convective boundary locations.

\subsection{Dependence on $\rmax$ and $\rmin$}
The value of $\rmin$ changes the depth of the resonant cavity where IGWs can propagate --  for $\rmin=0.6\rstar$ it is very shallow, spanning only about $0.12 \rstar \sim 1.4 H_p(\rschw)$, while for $\rmin=0.4\rstar$ it spans approximately $0.32\rstar \sim 3.9 H_p(\rschw)$.
As a result the frequency distribution of the $g$-modes changes significantly.
For instance, for the $\rmin=0.6\rstar$ simulations there are fewer $g$ modes at high frequencies than in the $\rmin=0.4\rstar$ simulations, as shown in \cref{fig:spec1d_l2_GYRE}.
Apart from this however, the power spectrum is quite similar between the two sets of simulations, particularly for $\omega \ga \SI{5}{\micro Hz}$.
This is consistent with the earlier observations that the value of $\rmin$ does not seem to have any significant effect on the depth of the overshooting layer and the convection zone, for the presented diagnostics at least (see \cref{sec: overshoot depth}).
This suggest that the detailed frequency distribution of the $g$-modes has no direct impact on the overshooting dynamics.

On the other hand, the value of $\rmax$ has no impact on the frequency distribution of the $g$-modes.
However, the amplitude of both travelling and standing waves in the radiative bulk grows significantly with increasing $\rmax$, as can be seen in both \cref{fig:spec2d,fig:spec1d_l2_GYRE}.
The qualitative relation can be intuitively expected because of the more vigorous convection in the higher $\rmax$ simulations which drives faster overshooting flows and a deeper overshooting layer.
\Cref{fig:spec2d} shows that the peak of the power spectrum in the radiative zone grows by approximately 3 orders of magnitude from the $\rmax=0.9\rstar$ to the $\rmax=0.99\rstar$ simulation, from $\SI{2.4}{}$ to $\SI{4.4e3}{cm^2 . s^{-2}}$ respectively.
At the same time, in the bulk of the convection zone the peak of the power spectrum grows by less than an order of magnitude (from
$\SI{5.8e6}{}$ to $\SI{3.0e7}{ cm^2 . s^{-2}}$) and
 so does $\sigma(\vur)$ (see \cref{fig:kinematics_cz}).
We note that $\sigma(\vur)$ when restricted only to the downflows in the convective bulk also does not have as strong $\rmax$ dependence as the IGWs amplitude (not shown).

With increasing $\rmax$ the energy of the IGWs is shifted consistently to higher frequencies (cf. \cref{fig:spec2d}).
This is consistent with theoretical expectations.
\Citet{lecoanet2013} link the frequency of the IGWs excited by  Reynolds stress at the convective boundary  with the characteristic convective frequency $\omega_{\conv} = 1/ \tau_{\conv}$, which grows with $\rmax$ (see \cref{tab:params}).
Similarly, \citet{pincon2016} link the frequency of the IGWs excited by penetrating plumes with the plume lifetime in the penetration zone.
While it is difficult to obtain a direct estimate of the lifetime of penetration events with Eulerian statistics, \citet{pincon2016} suggests the convective time scale $\tau_{\conv}$ as  a good first approximation.
Under this approximation, the plume lifetime would decrease with $\rmax$.
\footnote{
Alternatively, a simple ballistic approximation can be constructed from the average downward velocity at the Schwarzschild boundary, $\avg{v_\downarrow(\rschw)}_{\theta, t}$ and either of the penetration depths $l_{\bulk}^{\ke}$ and $l_{\max}^{\ke}$.
For the presented simulations, both $l_{\bulk}^{\ke}/\avg{v_\downarrow(\rschw)}_{\theta, t}$ and  $l_{\max}^{\ke}/\avg{v_\downarrow(\rschw)}_{\theta, t}$ decrease as $\rmax$ increases, supporting the notion that the plume lifetime decreases as $\rmax$ increases.}
Thus, both excitation mechanisms indicate that with increasing $\rmax$ more energy should be deposited in higher frequency IGWs.

In terms of length scales, at a given frequency, with increasing $\rmax$, the energy is shifted to larger length scales (smaller $\ell$).
This is also consistent with theory, since increasing $\rmax$ leads to convective eddies with larger length scale  (smaller $\ell_{\textrm{eddy}}$).
In that case, they are expected to excite waves with $\ell \leq \ell_{\textrm{eddy}} $ \citep{lecoanet2013}.

Overall, the dependence of the IGW power spectrum in the radiative zone on $\rmax$ are qualitatively analogous to the effect of luminosity boosting discussed by \citet{lesaux2021}, who also note an  increase in power and shift to higher frequencies and larger length scales with increasing luminosity.

\section {Summary and discussion}\label{sec:conclusion}

In this study, we investigate the effects of the radial extent on the dynamics of hydrodynamic 2D simulations of a solar-like stellar model.
We consider the dynamics in the convective, overshooting and radiative zones.

The location of the outer boundary plays a crucial role in determining the convective intensity as measured by the radial velocity and temperature fluctuations, \ie $\sigma(\vur)$ and $\sigma(T)$.
As $\rmax$ is increased, the steep hydrostatic density stratification leads to a decrease of the heat capacity of the outer layers and a corresponding increase in the temperature fluctuations.
The latter is required, so that the convective heat flux can transport the given fixed luminosity.
As we do not consider $\rmax$ high enough to include layers with inefficient convection, the radiative flux is negligible throughout the simulated convection zone.
The larger temperature fluctuations in the outer boundary layers drive faster radial flows through buoyancy.
In turn the faster radial flows advect  larger temperature fluctuations out of the diffusive thermal boundary layer.
These fluctuations then travel quasi-adiabatically in the convective bulk before dissipating at the lower convective boundary layer and driving convective boundary mixing.

The more intense convection leads to a deeper overshooting layer below the lower convective boundary.
We measure this by the radial kinetic and convective heat fluxes.
To characterise both the typical and extreme overshooting events we consider both the mean and the extreme overshooting depth, following \citet{baraffe2021}.
Both diagnostics grow rapidly with increasing $\rmax$.

Over several hundred $\tconv$ the overshooting leads to a small but measurable excess heating of the background in the overshooting layer.
The heating is well-correlated with strong mean background shear, indicating that the process responsible for it is likely mechanical mixing of adiabatically compressed hot convective material.
For a thermally relaxed state to be reached the excess heating must be balanced by a feedback mechanism, \eg naturally with the increased background temperature comes an increase of the radiative flux.
Despite the long simulation times (between $\sim 340 $ and $ 730\, \tconv$) a thermally relaxed state could not be reached in the simulations, indicating that the relaxation time of the excess heating is significantly longer (at least $O(10^3\, \tconv)$).

The location of the inner boundary of the simulations $\rmin$ is expected \emph{a priori} to change the properties of the internal gravity waves in the radiative zone.
In particular, it affects the frequency of the standing waves (the $g$-modes) and some wave reflection off the bottom boundary is observed for shallower radiative zones.
However, this appears to have no strong or consistent impact on the power spectrum of the travelling waves, the convection zone and overshooting layers.
This insensitivity is expected to hold while the feedback of the IGWs on the overshooting layer is weak, \eg the radiative zone is deep enough to allow for significant wave damping and no direct interaction between the overshooting  and bottom boundary layers.

The position of the outer convective boundary impacts significantly the internal gravity waves in the radiative zone.
Even though the frequency dependence of the $g$-modes remains unaffected,  the energy of all IGWs is increased
because of larger overshooting velocities (which also imply a larger overshooting depth).
In particular, larger-frequency waves are more strongly excited in agreement with theoretical expectations \citep{lecoanet2013}.

\subsection{Luminosity boosting analogy}
It is interesting to note that the dependence of the overshooting layer on $\rmax$  is qualitatively similar to that on luminosity boosting discussed by \citet{baraffe2021}.
Luminosity boosting is an often-used technique in stellar hydrodynamics, which shortens the thermal time-scale of the model and brings it closer to the convective one.
This leads to a substantial reduction in computational costs and makes reaching thermal equilibrium feasible given a large enough boost factor.
However, it has to be done carefully in order to maintain the original background stratification of the un-boosted model,
\eg the thermal diffusivity has to be increased proportionally  and stratification in the convection zone has to be adjusted to as close to adiabatic as numerically feasible.
Even then, as discussed in \cite{baraffe2021}, it comes at the cost of affecting the overshooting dynamics by increasing the overshooting depth, the local heating in the overshooting layer and the shape of the IGW spectrum \citep{lecoanet2019a}.

This effect is qualitatively similar to the dependence we find on $\rmax$, especially considering that the stellar models used by \citet{baraffe2021} and in this study have significantly different background stratification and luminosity.
Recalling that all the presented simulations have the same luminosity, this highlights that the underlying dependency is not on the energy flux \emph{per se}, but on the convective intensity, in the form of e.g. the temperature contrast and the radial velocity.
Ultimately, because of the adiabatic evolution in the convective bulk, the dependence is on the stratification at the outer boundary, which determines the fluctuations required to transport the luminosity there.

It is difficult to make a direct quantitive comparison between the effects of increasing $\rmax$ in the simulations presented here and the effects of boosting the luminosity discussed by \citet{baraffe2021}, because the background stellar models are significantly different.
(For instance the boosted simulations have convection zone with much smaller super-adiabaticity $\sim 10^{-8}$, deeper convective envelope $\rschw \sim 0.68 \rstar$, smaller stellar radius $\rstar \sim \SI{0.8}{ \rsun}$ and larger stellar luminosity).
However, as an indication we note that the difference in the overshooting depths between the $\rmax=0.9\rstar$ and $\rmax=0.99\rstar$ simulations are comparable to the effect of boosting the luminosity by a factor of 100 in the boosted simulations.
Note that \Citet{baraffe2021} caution against using significantly larger boost factors in any case, as they may lead to evolution of the background model and spurious changes of \eg the IGW spectrum.
Similarly, extrapolating the presented results to values of $\rmax$ significantly above the $0.99\rstar$ level is challenging, because this  would include surface layers where the opacity quickly decreases and additional physics would need to be considered
(\eg the increasing impact of radiative cooling requires a detailed treatment of radiative transport).

\subsection{Outlook }

Because of the steep stratification it is computationally challenging
to perform reasonably well-resolved global stellar hydrodynamic simulations with $\rmax= 1 \rstar$
with realistic radiation physics and self-consistent surface boundary layers.
Hence quantitative predictions of convective velocities, IGW spectra and overshooting depths should always be considered with great care.
This is a big part of the challenge of solving the so-called solar convective conundrum \citep{hanasoge2012,schumacher2020,hotta2021,vasil2021} and is the reason to refrain from a more quantitative analysis of the data.
As $\rmax$ is increased towards unity the time and length scales of the boundary layers decrease
while the cooling efficiency increase.
Eventually, an asymptotic limit must be reached for the amplitude of the radial velocity and temperature fluctuations in the convection zone (and consequently for the overshooting depth and IGW power spectrum).
However, it would be computationally unfeasible to seek such a limit with the uniform radial grids used in this study.
Encouragingly, \citet{hotta2019} find similar profiles of the root-mean-square velocities in local convection zone simulations truncated at $\rmax=0.992 \rstar$ and $\rmax=1\rstar$, when the former is augmented with an artificial surface cooling layer.
They attribute the similarity of the results to efficient mixing in the near-surface region, which reduces the non-locality of the convection.
This indicates that it may be possible to model the influence of the truncated, and difficult to resolve, surface boundary layers.
Further studies are needed to examine the influence that such modelling may have on the overshooting depth and IGWs.
It also remains to be established how the presented non-local dependencies translate to 3D convection and how they interact with other important processes like rotation and magnetism, for which a follow-up study is currently under way.
However, the qualitative trends this study has highlighted (the significant dependence on $\rmax$ and the weaker dependence on $\rmin$) should be robust results that can be linked to physical mechanisms which operate in the stellar hydrodynamics context.


\section*{Acknowledgements}
This work is supported by the ERC grant No. 787361-COBOM
and the consolidated STFC grant ST/R000395/1. The authors
would like to acknowledge the use of the University
of Exeter High-Performance Computing (HPC) facility ISCA
and of the DiRAC Data Intensive service at Leicester, operated
by the University of Leicester IT Services, which
forms part of the STFC DiRAC HPC Facility. The equipment
was funded by BEIS capital funding via STFC capital
grants ST/K000373/1 and ST/R002363/1 and STFC DiRAC
Operations grant ST/R001014/1. DiRAC is part of the National
e-Infrastructure.


\section*{Data Availability}
The data underlying this article will be shared on reasonable request to the corresponding author.



\bibliography{DepthStudyArticle} 
\bibliographystyle{mnras}

\appendix

\section{Convective heat flux derivation}
 \label{app:h_flux}
In this work we use \cref{eq:flux_h ave} as a definition of the convective heat flux.
This is derived from the angular average of the total energy density budget
 \begin{align}
	\pdt \avg{\etot}_\theta &= -\pdr \avg{\half \rho \vur \vu^2  - \chi \grad T +  \rho \vur h + \rho \vur \Phi}_\theta \\
	 & = -\pdr \avg{\v{F}^{\ke}_r  + \v{F}^{\chi}_r +  \v{F}^{\conv}_r + \v{F}^{\bkgr}_r}_\theta,
 \end{align}
 where we introduce $\etot$ as the sum of the kinetic, gravitational and internal energy densities,
 $ \v{F}^{\chi}_r  = -\chi \grad T$ as the radiative flux and
 $ \v{F}^{\bkgr}_r = \avg{\rho \vur}_\theta  \avg{\Phi + h}_\theta$ as the flux due to the mean stratification.
We recall that $\v{F}^{\conv}_r = (\rho \vu)' h' $ and $'$ designates deviations away from the instantaneous angular average.
In principle, given that in the convection zone $\avg h \gg h'$, the contribution of the mean specific enthalpy to the energy transport may be non-negligible in the not-perfectly-relaxed case even with a  small mean radial momentum.
However, this is largely compensated by the hydrostatic stratification through the mean potential energy $\avg{\Phi}$.
Using the \emph{fundamental thermodynamic relation} $\di h = T \di s + \rho^{-1} \di p$ (with $s$ the specific entropy) we can express $h'' = \avg{h+ \Phi}_\theta$, the mean enthalpy excess away from hydrostatic equilibrium (HSE), as
\begin{align}
	\pdr \avg{h''}_\theta &= \avg{T \grad s+ \rho^{-1}\grad p - \rho^{-1} \grad p_{HSE}}_\theta\\
	&=  \avg{T \grad s+ \rho^{-1}\grad p''}_\theta,
\end{align}
where we define  $\grad p_{HSE} = - \rho \grad \Phi$ as the pressure gradient in HSE and   $p''= p - p _{HSE}$ as the pressure perturbations away from it.
This illustrates that mean enthalpy flux contains  pressure perturbations with respect to HSE  and entropy perturbations.
Given a dynamical time scale of $10^3\,\si{s} - 10^4\,\si{s}$ in the convection zone (depending on radius), any deviations of the mean stratification from HSE are quickly corrected, so the $p''$ term can be neglected over the time scales considered in this study.
The mean entropy perturbations are at the order of $10^{-5}$ and are therefore similarly negligible.
This justifies neglecting the background flux $ \v{F}^{bkgr}_r$ and considering $\v{F}^{\conv}_r$ as the convective heat flux.

\end{document}